# Alteration of skeletal muscle energy metabolism assessed by 31P MRS in clinical routine, part 1: Advanced Quality Control pipeline

Antoine Naëgel[1,2], Hélène Ratiney[1], Jabrane Karkouri[1,2,3], Djahid Kennouche[1,4], Nicolas Royer[1,4], Jill M. Slade[5], Jérôme Morel[6], Pierre Croisille[1,7], Magalie Viallon[1,7]

[1] Université de Lyon, INSA-Lyon, Université Claude Bernard Lyon 1, UJM-Saint Etienne, CNRS, Inserm, CREATIS UMR 5220, U1206, Lyon, France, [2] Siemens Healthcare SAS, Saint-Denis, France, [3] Wolfson Brain Imaging Center, University of Cambridge, Cambridge, United Kingdom, [4] LIBM - Laboratoire Interuniversitaire de Biologie de la Motricité, [5] Department of Radiology, Michigan State University, East Lansing, USA ; [6] anaesthetics and intensive care department, UJM-Saint-Etienne, Centre Hospitalier Universitaire de Saint-Étienne, Saint-Etienne, France ; [7] Radiology department, UJM-Saint-Etienne, Centre Hospitalier Universitaire de Saint-Étienne, Saint-Etienne, France

Abbreviations: ATP, adenosine triphosphate; BMI, body mass index; CV, coefficient of variation; $k_{PCr}$, rate constant of post-exercise PCr recovery; MS, multiple sclerosis; MVC, maximum voluntary contraction; PCr, phosphocreatine; Pi, inorganic phosphates; QCS, quality control score; QUEST, QUantitation based on QUantum ESTimation; $R_{ind}$, individual correction factor; $R_{fixe}$, fixed correction factor; $\tau_{PCr}Rec$, time constant of post-exercise PCr recovery; $\tau_{Pi}Rec$, time constant of post-exercise Pi recovery; $\tau_{PCr}Ex$, time constant of exercise PCr depletion; $\tau_{Pi}Ex$, time constant of exercise Pi depletion;

Corresponding author: Magalie Viallon, Université de Lyon, INSA-Lyon, Université Claude Bernard Lyon 1, UJM-Saint Etienne, CNRS, Inserm, CREATIS UMR 5220, U1206, Lyon, France.

Email : magalie.viallon@creatis.insa-lyon.fr

**Abstract Summary**

Background: Implementing a standardized $^{31}$P-MRS dynamic acquisition protocol to evaluate skeletal muscle energy metabolism and monitor muscle fatigability[1,2], while being compatible with various longitudinal clinical studies on diversified patient cohorts, requires a high level of technicality and expertise. Furthermore, processing data to obtain reliable results also demands a great degree of expertise from the operator. In this two-part article, we present an advanced quality control approach for data acquired using a dynamic 31P-MRS protocol. The aim is to provide decision support to the operator in order to assist in data processing and obtain reliable results based on objective criteria. We present first in part one, an advanced data quality control (QC) approach of a dynamic $^{31}$P-MRS protocol. Part two is an impact study demonstrating the added value of the QC approach to explore clinical results derived from two patient populations with significant fatigue: COVID19 and multiple sclerosis (MS).

Experimental: $^{31}$P-MRS was performed on a 3T clinical MRI in 175 subjects from clinical and healthy control populations conducted in a University Hospital. An advanced data QC Score (QCS) was developed using multiple objective criteria. The criteria were based on current recommendations from the literature enriched by new proposals based on clinical experience. The QCS was designed to indicate valid and corrupt data and guide necessary objective data editing to extract as much valid physiological data as possible. Dynamic acquisitions using an MR-compatible ergometer ran over a rest(40s), exercise(2min), and a recovery phase(6min).

Results: Using QCS enabled rapid identification of subjects with data anomalies allowing the user to correct the data series or reject them partially or entirely as well as identify fully valid datasets. Overall, the use of the QCS resulted in the automatic classification of 45% of the subjects including 58 participants that had data with no criterion violation and 21 participants with violations that resulted in the rejection of all dynamic data. The remaining datasets were inspected manually with guidance allowing acceptance of full datasets from an additional 80 participants and recovery phase data from an additional 16 subjects. Overall, more anomalies occurred with patient data (35% of datasets) compared to healthy controls (15% of datasets).

Conclusion: This paper describes typical difficulties encountered during the dynamic acquisition of 31P-MRS. Based on these observations, a standardized data quality control pipeline was created and implemented in both healthy and patient populations. The QC scoring ensures a standardized data rejection procedure and rigorous objective analysis of dynamic 31P-MRS data obtained from patients. The contribution of this methodology contributes to efforts made to standardize the practices of the $^{31}$P-MRS that has been underway for a decade, with the ultimate goal of making it an empowered tool for clinical research.

**Introduction**

Evaluating skeletal muscle energetics and metabolism is of clinical interest to monitor neuromuscular or cardiovascular degenerative diseases and quantify muscle fatigability[1–3]. $^{31}$P-MRS is a non-invasive technique of choice for dynamically assessing the concentration of phosphorylated metabolites, which are directly related to the respiratory capacity of mitochondria [4–6].

However, deploying this technique in translational research or clinical context is not straightforward. $^{31}$P-MRS has been the subject of numerous studies that have pushed forward the standardization of protocols while questioning the entire protocol steps, from acquisition to data processing. Concerning data acquisition, one can quote Roussel's study[7] on the variability between subjects under different exercise conditions and Ratkevicius' study[8] evaluating the impact of voluntary versus electrical muscle contraction. When considering data analysis, an important work from Layec et al.[9] proposed to study the impact of different models of respiratory control on evaluating the rate of maximal mitochondrial adenosine triphosphate (ATP) synthesis.

The reproducibility and reliability of the dynamic $^{31}$P-MRS measurements have been widely published. These include studies on metabolic markers[10], specialized cohorts[11] (trained and sedentary), various types of exercise[12] (moderate and high intensity), and multi-site studies with different ergometers[13]. Overall, these studies demonstrate good reproducibility and reliability, allowing the use of $^{31}$P-MRS dynamic measurements in longitudinal studies and supporting the evidence base for introduction into the clinic.

Most importantly, a recent expert consensus on $^{31}$P-MRS in skeletal muscle has provided general recommendations on the use and operation of this technique[4]. Nevertheless, translating $^{31}$P-MRS in a clinical routine on large cohorts leads to additional challenges, lower compliance, and compromise hence asking questions on the necessity of more precaution or corrections during the analysis. The inclusion of numerous patients and the need for rapid but rigorous analysis of all datasets highlighted the need for a specific data quality control pipeline, based on precise indicators quantifying the deviation from perfect data acquisition.

The overall goal of part one is to improve dynamic $^{31}$P-MRS data analysis for application in populations with reduced health. The first aim is to provide a descriptive analysis of typically encountered problems resulting from reduced data quality or artefacts occurring during dynamic $^{31}$P-MRS acquisition. The second aim is to create and implement an advanced data quality control pipeline designed to ensure unbiased and accurate results. The third aim is to apply the quality control pipeline to score datasets using quality metrics to guide decisions on acceptable quality including strategies to retain partial datasets. The final aim is to determine if particular groups (patient groups) are more vulnerable to corrupted data, e.g. those that would most benefit from a robust quality control approach.

**Experimental**

*Studied populations*

This work was conducted on 175 subjects in a University Hospital, from 21.11.2017 to 29.04.2022, including healthy subjects and clinical patients with MS or COVID19. All subjects provided written informed consent and were recruited in protocols approved by the Institutional Review Boards (CPP Nord Ouest VI, # 19.02.22.52507, and CPP Ile de France VIII, #20 04 05), in agreement with the principles in the Declaration of Helsinki, and/or registered at ClinicalTrials.gov (NCT04363606, NCT05031598). The 175 subjects included 19 COVID19 patients (age: 64±13 years, Body Mass Index (BMI): 25.98±4.28 kg/m$^2$), 38 MS patients (age: 45±9 years, BMI: 25.26±4.58 kg/m$^2$), and 118 healthy subjects (age: 47±12 years, BMI: 24.73±3.67 kg/m$^2$), demographic data are provided in Table 1 of the supplementary material.

*MRS set-up and resting muscle assessments*

Spectroscopy and NMR imaging were performed on a 3T MR unit (MAGNETOM PRISMA, Siemens Healthineers, Germany). Subjects were lying supine with a $^1$H/$^{31}$P dual tuned transmit/receive surface coil (Circular with a radius of 8 cm and a penetration depth of about one radius, Rapid GmbH, Würzburg, Germany) positioned under the calf muscle and firmly attached to the leg. Extraneous movements of the body were minimized using a strap across the abdomen and the instruction to keep hands on the thorax. The participant's dominant foot was placed in an MR-compatible plantar flexion ergometer (ErgoSpect, Innsbruck, Austria) equipped with the Trispect module.

A non-localized MRS-FID sequence was used to acquire data; the sequence included the application of saturation bands obtained with adiabatic pulses to minimize signals from unstressed muscles and bones[14]. The coil position was chosen to cover and focus on the medial part of the gastrocnemius and the soleus muscles, while the saturation band was placed on the anterior and lateral leg compartments (see Supplementary Material Figure 1).

Two resting $^{31}$P-MRS acquisitions were performed before the exercise, with TR=30sec (12 acquisitions) and TR=4sec (32 acquisitions) to obtain a fully-relaxed and a $T_1$ saturated spectrum of the quiescent muscles[15] respectively. The short TR sequence is essential for dynamic exercise studies using $^{31}$P-MRS.

*Dynamic MRS protocol and assessment of muscle force*

After the resting acquisitions, each participant was asked to practice the exercise protocol before any dynamic/exercise MRS data were acquired to familiarize themselves with the timing and expected tasks. Patients were trained to properly engage the target muscle groups (planter flexors) by guidance from the lead clinician. This included verbal remarks such as "push through the ball of the foot" and "push as if you are doing a calf raise" and to avoid using the hips for the exercise. Additionally, a visual check over the lower limb was done watching the ankle/calf for the expected contraction as well as watching for inappropriate recruitment of the thigh and lifting or twisting of the pelvis. Proper recruitment was ultimately checked during the QC; the QC established a minimum change in PCr, and this serves as a strong indicator of whether the exercise was done with the target muscle group, albeit after the fact. The subjects were provided with real-time visual feedback of their performance on an MR-compatible screen mounted ~2m directly in front of their line of vision. They received clear, standardized visual instructions asking for periodic plantarflexion contractions on the ergometer every 4s (Visual stimuli were displayed using E-Prime (version 2.0.10.261), Psychological Software Tools (Sharpsburg, PA, USA)). Next, three maximal isometric voluntary contractions (MVC) were recorded; the participants asked to press the pedal as much as possible. Each MVC was interspersed with at least 60s rest, and force output was recorded.

Following the MVC assessment and five minutes of resting recovery time, the dynamic MRS protocol began. Participants were asked to push as hard as possible (performing isometric MVC) during the MRS acquisition while following the visual instructions on when to contract. The dynamic MRS acquisitions ran over a 40s rest phase (10 acquisitions), a 2min exercise phase (30 acquisitions), and a final 6min recovery and rest phase (90 acquisitions). This exercise was designed to elicit a moderate intensity with 1s MVC of the ankle performed every 4 sec (0.25 Hz). The timing of the contraction was synchronized to occur prior to each $^{31}$P dynamic MRS acquisition to acquire spectra when the calf was in a neutral position. The selection of a 4s TR was chosen as an optimal compromise ensuring optimal $T_1$ signal recovery and SNR, motion-free spectral acquisition, and short enough temporal resolution to warranty appropriate monitoring of the dynamic processes and physiological changes of interest. This TR duration (4s) also allows the patients to see and follow the instruction visually, react to it by performing a single push, and then return to the neutral resting position, and to attain motion-free valid spectra during the exercise. This exercise intensity also ensures steady-state conditions are achieved i.e., the absence of net changes in metabolite concentrations will be met[16] while avoiding muscle acidosis. The force output data (N.m$^{-1}$) were recorded throughout the exercise and were analyzed to determine the amplitude for contraction exerted during exercise, expressed as a percentage of initial MVC.

*³¹P-MRS Quantification*

The data were processed with MATLAB and the QUEST (QUantitation based on QUantum ESTimation) method in its command-line version [17]. After a manual data phasing procedure using the PhaseTool GUI [18] and a 5 Hz exponential apodization, the QUEST method used a metabolite basis set consisting of PCr, Pi, α-, β-, and γ-ATP for which amplitudes, frequency shifts, and damping parameters were estimated.

The metabolite basis set was composed of Lorentzian line shapes with 47 Hz of linewidth and at the resonance positions of distinct signals related to PCr (0ppm), Pi (5.02ppm), ATP-β (-16.26ppm) and a doublet for γ (-2.48ppm) and α (-7.52ppm) ATP. The ATP-β was approximated to a singlet for the acquisitions with TR=4s due to low SNR, therefore the same process was consistent for the acquisition with TR=30s.

Flexible constraints were introduced in QUEST regarding the damping factor and frequency offset. The lower and upper bound for extra-damping and frequency shift is [-40 40] Hz and [-30 30] Hz, respectively. These constraints enable flexibility during the estimation procedure to follow the frequency and damping fluctuations generated during the muscle exercise. The evaluation of the millimolar concentration of phosphorus metabolites was based on the standard assumption that [ATP] is 8.2 mmol/L of cell water [5]. The concentration and frequency of the metabolites PCr, Pi, and ATP were measured and extracted throughout the three sequential phases of the dynamic protocol: rest, exercise, and recovery (see Supplementary Material Figure 1). Muscle pH was calculated from the frequency difference between PCr and Pi, as in [4,19]:

$$pH = pKa + \log\left(\frac{\delta - pKam}{pKaM - \delta}\right) \ (Eq2)$$

With pKa = 6.75, pKam = 3.27, pKaM = 5.69, and δ the chemical shift between Pi and PCr. The pH was averaged at rest to obtain the resting pH and just after exercise to obtain the post-exercise pH.

We also examined PCr depletion (percentage) during exercise:

$$\text{PCr depletion} = \left(1 - \frac{[\text{PCr}]_{\text{Post}}}{[\text{PCr}]_{\text{Rest}}}\right) \cdot 100 \ (Eq7)$$

Where [PCr]$_{\text{Post}}$ reflects the PCr at the end of the exercise phase. A single-exponential fit was performed on the exercise and recovery period of PCr and Pi amplitudes, to extract the time constants of PCr ($\tau_{\text{PCr}}$Ex and $\tau_{\text{PCr}}$Rec) and Pi ($\tau_{\text{Pi}}$Ex and $\tau_{\text{Pi}}$Rec) such that:

$$PCr(t) = PCr_b \pm \Delta PCr \cdot \left(1 - e^{-t/\tau_{\text{PCr}}}\right) (Eq8)$$

$$Pi(t) = Pi_b \pm \Delta Pi \cdot \left(1 - e^{-t/\tau_{\text{Pi}}}\right) (Eq9)$$

Where PCr(t) or Pi(t) is the value of PCr or Pi at time t, PCr$_b$ or Pi$_b$ is the baseline value of PCr or Pi. For the exercise phase fit, the baseline value is the value at rest and for the recovery phase fit, the baseline value is [PCr] and [Pi] post-exercise value. ΔPCr and ΔPi represent the difference between the baseline and steady-state exercise values. Note that $\tau_{\text{PCr}}$Rec and $\tau_{\text{Pi}}$Rec are specific to the recovery phase, although they can be influenced by the exercise phase, as described in the next section.

*Adaptive Analysis and Quality Control*

Despite appropriate design including standardized motion, visual instructions, and verbal encouragements, sources of data corruption may remain. These observations include a tendency to anticipate the push resulting in early and poorly timed pushes. Performance may be impacted by slow contractions associated with a gradual increase in pressure on the pedal exceeding the expected contraction window. Conversely, a loss in the amplitude of the push due to motivation issues (low patient effort), muscle recruitment issues (improper exercise) or muscle fatigue were observed. When pushing amplitude is greatly reduced, this can lead to limited metabolite changes confounding interpretation of behavior data. Additionally, recovery data may be corrupted by a failure to completely stop exercise at the end of the cued contractions. The latter deserves special attention since the monoexponential adjustment, used to extract the time constants of PCr and Pi during the exercise and recovery phase, strongly depends on the last point of the exercise, i.e., the first point for the recovery period to be considered. Therefore, based on the literature[4] and our own experience (processing hundreds of datasets across various populations[20–25]), we proposed a quality control score (QCS), intended to reflect the quality of the data acquisition and of the estimated parameters as described in Figure 1.

The QCS pipeline works by entering MRS post-processed spectra along with the standardized timing of the phases (rest, exercise, recovery). The analysis includes several criterion checkpoints and each one is associated with a score. The score guides decisions about the acceptance of dynamic/behavior measures. This includes next steps for datasets that need manual inspection and objective guidance for correction and data editing.

QCS is based on the following criteria:

- PCr Depletion < 20%: If the percentage of PCr depletion between resting and post-exercise levels is less than 20%, the interpretation of τPCr recovery is not valid[4].
- R2 $\tau_{PCr}$Rec |$\tau_{Pi}$Rec < 70%: The coefficient of determination R2 was used to evaluate the coherence between the mono-exponential model and the data of each subject, an R2 < 70% reflects poor data quality.
- Outliers identification based on the time derivative of (PCr+Pi): See explanation below.
- $\tau_{PCr}$Ex| $\tau_{Pi}$Ex > 100s: the adjustment on the exercise part reflects the kinetics/dynamics of the metabolites, and the extracted time constants > 100s reflect a linear behaviour unlikely to represent mitochondrial function which is widely accepted as exponential.
- R2 $\tau_{PCr}$Ex|$\tau_{Pi}$Ex < 70%: same as the R2 for the recovery part.
- Coefficient of Variation (CV) of $\tau_{PCr}$Rec|$\tau_{Pi}$Rec > 10%: See explanation below.

The outlier indicator may be explained in detail as follows: theoretically, during muscular exercise, the decrease of PCr concentration is coupled with an increase of Pi concentration such that (PCr+Pi) is constant. Under this assumption, a large variation in (PCr+Pi) reflects corrupted data during exercise[16]. As an indicator of this behaviour, the time derivatives of the sum of (PCr+Pi), PCr and Pi concentration are used. Outliers were identified as deviating from the mean by more than three standard deviations.

The quality criteria are weighted according to their importance. The quality criteria essential for the extraction of physiological parameters related to the recovery phase are weighted strongly with a -3 (PCr hydrolysis and R2 $\tau_{PCr}Rec\,|\,\tau_{Pi}Rec$) with direct support from the consensus paper[4,12]. Meanwhile the quality criteria related to the exercise phase are weighted less (0.5 – 1.0) to detect corrupted data resulting from poorly timed contractions or movement during acquisition; these datasets will require manual inspection and possible corrections to accept the results (see Figure 1). These lower weightings allow multiple violations during the exercise to be considered and evaluated to assist with retention of as much valid data as possible.

This weighting results then in a score that reflects the associated criteria. The different thresholds and intervals associated with decision making are therefore set to exclude data presenting a criterion associated with a high weighting. The data exclusion threshold is therefore set at -3 in our case and indicates invalid interpretation of exercise and recovery data (data are automatically excluded) with no further pathway to re-evaluate the behavior data. Data with a QCS between 0 and -3 are directed to a manual analysis with 2 levels of decision support. Data with a QCS of 0 is considered acceptable.

Using objective criteria, the QCS enables the operator to decide whether to accept the data, manually inspect them, or exclude them. It is worth noting that data corruption can affect the exercise phase, the recovery phase only, or both phases. This scoring is a valuable step to ensure robust data while saving time that otherwise would be required to manually assess data on each of the criterion. The behavior/dynamic results from scores of 0 and < -3 will not need further review and therefore the results after QCS therefore will be automatically accepted.

If outliers were detected by the time derivative of the PCr and Pi indicator, manual inspection was considered mandatory to validate the first point of the recovery and the corresponding parametric adjustments. To identify valid initial spectra/datapoint and inspect the ultimate impact of this reselection, we propose an iterative adjustment procedure as follows. The first, second, and third points of recovery, respectively, at times 160, 164, and 168s of the protocol were sequentially excluded, and the corresponding four monoexponential fits for the recovery phase and the two metabolites (PCr and Pi) were obtained. A Coefficient of Variation (CV) between the estimated time constants for the recovery phase was then computed such as:

$$CV = \frac{\sigma}{\mu}$$

Where $\sigma$ and $\mu$ are the standard deviation and mean of the four estimated time constants for the recovery phase. Hence, CV stands as an additional indicator to appreciate the behavior of the exercise-dependent adjustment.

Finally, the SNR and the FWHM, described in the consensus paper[26], of the PCr and Pi peaks for the long and short TR acquisitions were also explored (see Supplementary Material Table 2). Concerning the short TR acquisitions, these indicators are taken at the resting and post-exercise phases. These indicators give a good appreciation of the quality of the spectra and are complementary to our quality control analysis of the dynamic experiment.

**Results**

Figure 2 illustrates subjects having a QCS between 0 and -7.25 representing high quality data as well as several compromised datasets targeted by the proposed QC pipeline. The examination of these different behaviors revealed by the QCS highlights the impact of each criterion contained in the score. A QCS of less than -3 indicates automatic exclusion of the patient, this score being attributed largely to the criterion of PCr depletion (PCr depletion < 20%) or by the coefficient of determination R2 less than 70, representative of poor data quality of τPCrRec or τPiRec. Figure 2F shows a subject with a QCS < -3, with PCr depletion less than 20% and R2 less than 70. In this case, the dynamic data during exercise and recovery are not valid for estimating behaviours.

A QCS between -1 and -3 requires precise indicators to evaluate the exercise and recovery part. The Outliers indicator, described in the previous section, is very sensitive to the signal variation during the protocol and coupled with the quality indicators relative to the adjustment of the exercise part (τPCrEx or τPiEx > 100s and R2 τPCrEx or τPiEx < 70%) and to the CV of the τPCrRec or τPiRec, they highlight problems that occurred during the protocol in a more refined manner.

An outlier detected during the exercise and a high CV of the PCr adjustment indicates corrupted data due to an exacerbated movement during the exercise which resulted in signal loss impacting both exercise metabolites and the first points of the recovery, see Figure 2D. The QCS of -2.25 leads to a manual inspection with a strong marker to inspect the transition between exercise and recovery by evaluating the first spectra associated with the start of the recovery period. Figure 3 shows the amplitude of metabolites during the protocol and their derivative for a subject with acceptable data (QCS = 0, same subject as Figure 2A) and in the case of a subject with corrupted data (QCS = -2.25) as presented in Figure 2D. Figure 4 provides stack plots of PCr and Pi for these same subjects.

Figure 5 shows the iterative adjustment procedure of the PCr, during the recovery phase, used in the case of subjects with corruption in τPCrRec|τPiRec by the exercise phase. The figure shows three different types of cases with QCS = 0 (the same subject shown in Figures 2A, 3, 4), QCS = -2.25 (the same subject shown in Figures 2D, 3 and 4) and QCS = -7.25 (the same subject shown in Figure 2F). For the case of QCS = -2.25, the adjustment selected (corresponding to the best R2 and <70%) is the one with the exclusion of the first point (Figure 5, middle).

The application of the QCS on the 175 subjects is depicted in Figure 6. The pie charts detail the results of the QCS and manual inspection each for group of subjects. Overall, the outcome yielded 33% of the subjects with a QCS = 0 with high quality data throughout and all data are automatically included. Meanwhile 55% of subjects have a QCS between 0 and -3 and are thus need manual inspected and 12% have a QCS <-3 and are thus automatically excluded. Therefore, the use of the QCS resulted in the automatic classification of 45% of the subjects. After manual inspection of the data with a QCS between 0 and -3 and adjustment to the fitting point when needed, the full acquisition is valid for 83% of the subjects and only the recovery behavior for the rest (17%). The portion of inclusion and exclusion of data is different between the healthy and clinical groups; 85% of the healthy sample had valid exercise and recovery data compared to 68% and 58% for the patients with COVID19 and MS, respectively. The percentage of complete data exclusion in the clinical groups is 2.76 times higher than in the healthy group; 21% for each patient group compared to 8% for the healthy control group.

To ensure that our dynamic parameters, such as $\tau PCr$, are meaningful, it is necessary to ensure 1) significant depletion of PCr and 2) absence of muscle acidosis. As expected from the mild exercise design, there was no significant change in pH during exercise nor severe acidosis detected in any participant (see Supplementary Material Table 3). The muscular exercise led to a significant depletion of PCr (40%, n =138). These checks attest that the data have been acquired under favorable conditions and can support the analysis and the results.

The orders of magnitude of the SNR and FWHM are in agreement with the expectations concerning the quality of the spectra. This ensures high spectral quality with the selected parameters for the protocol, see the results in Table 2 attached in the supplementary material.

**Discussion**

We propose a quality control pipeline including a data scoring (QCS and rating scale) to ensure data selection for subsequent analyses because data may be corrupted by a critical dampening of the PCr and Pi peaks occurring during exercise, as illustrated in the cohorts. Large signal fluctuations primarily result from motion during the exercise. However, rapid signal fluctuations in the data may also arise during other phases of acquisition, when the subject does not remain relaxed or has significant involuntary muscle twitching. Indeed, a near complete loss of the total high energy phosphate signals was observed in some subjects during exercise contractions associated with improperly timed contractions that alter the loading of the coil and thereby reduce the signal. Despite selecting an exercise protocol design that allowed adequate time between contractions and the synchronized exercise and spectral acquisition, the data were compromised for many participants, particularly for patients. This means that despite careful protocol design and patient preparation, data can still be compromised, underscoring the importance of objective and systematic quality control measures during analysis.

When the same data quality control pipeline was applied to all subjects, we identified four subjects requiring a reselection of the first time point for subsequent fitting of the metabolite recoveries. In our scenario of a mild exercise level, the reselection of the starting point of the recovery fit is not anticipated to introduce a significant bias since a monoexponential pattern is expected. The monoexponential PCr recovery pattern cannot be generalized to all exercise protocols. It should be considered an oversimplification when the initial rate of PCr resynthesis follows an exercise scheme including a glycolytic component[27,28]. For example, data from the dorsiflexors following a brief burst exercise (0.50 Hz) showed that removal of the first points of recovery prolonged the PCr time constant[29]. However, the selection of a highly oxidative muscle group (plantar flexors), the mild exercise intensity of the current study (0.25 Hz), and similar exercise pH levels between groups that remained at or above resting levels reduce the likelihood of bias. Moreover, the dorsiflexors have low oxidative capacity, despite being comprised of 75% slow myosin ATPase[30]. Nevertheless, it could be argued that the initial rate of PCr resynthesis glycolytic component could be physiology dependent and that even this mild exercise for the plantar flexors could have a different glycolytic component between the patient and control. Indeed, the monoexponential pattern of phosphocreatine recovery after muscle exercise is a particular case of more complex behavior and could appear as an over-simplification, especially in patients[27]. Nevertheless, according to the consensus paper[4], methods using bi-exponential functions to investigate and extract the 'early-recovery' component are not definitive and may be of interest in the presence of a significant change in pH, which is not our case. The depletion of PCr reached the level of expectation to create a significantly detectable change while preserving the pH[4,7], warrantying a fair comparison of $\tau_{PCr}$ and $\tau_{Pi}$ between the studied populations.

These results seem to confirm the strategic importance of systematically integrating objective data quality control when working in a clinical setting. The real impact on clinical outcomes will be assessed in Part 2. The introduced data quality control method allowed us to detect the corrupted data efficiently hence proposing an objective QCS procedure which resulted in limiting data exclusion, guaranteeing the maximum exploitation of the data by considering the exercise and recovery parts separately.

The development of additional quality indicators would allow comparisons of the QCS method to ascertain its pertinence and its expression on patient cohorts. Ultimately, the QCS method may also help to objectively quantify the effect of specific refinements in the protocol, such as the importance of pre-training sessions across challenging populations (seniors, children, patients). In light of the above-mentioned QC results, implementing a data quality control methodology can be used to strengthen the derived clinical parameters and final results. Based on recommendations from the literature and our own experience, the QC indicators and QCS complement the recommendations made in the [31]P-MRS consensus papers in the context of studies in clinical environments. The knowledge and technical expertise related to acquiring and exploiting [31]P-MRS data must be transcribed into explicit indicators of data quality. In this study, in addition to the physiological and clinical results described in the second part, the intention was to transfer a knowledge base from the research to the clinical environment to ensure the interpretation of physiological phenomena by [31]P-MRS. As a future development, the implementation of a real-time QCS could quickly lead the operator to repeat the measurement or pre-training session.

As mentioned before, quality control should exclude as few patients as possible from the study in a non-subjective way, preserving as much of the data as possible. This implies creating quality indicators that are as accurate as possible, targeting only corrupted data while preserving any true drifts associated with pathologies or treatments. While the QCS in the current study was applied across three distinct populations (COVID19, MS and healthy controls), it is expected that the QCS can also be used when monitoring treatment effects and across other pathologies. In this regard, the indicators should continue to be applied, enriched, and validated by the expert community for transparent and open research.

**Conclusion**

In this first part, the motivation and the impact of a specifically designed quality control were presented, demonstrating its capacity to target corrupted datasets, alert and guide the supervised analysis, and standardize the inclusion of patients from large cohorts. The typical problems encountered during acquisition affecting the processing of dynamic 31P-MRS data were described. The implementation of the Quality Control Score provided the opportunity to combine the literature recommendations and the operator's experience using objective exclusion criteria of subjects with corrupted data, in the context of mandatory transparent and open research. The application of QCS has shown considerable effects on the classification of acquired data and the extraction of subsets of data corresponding to specific phases of the acquisition protocol. Among our subjects, patients showed a higher proportion of corrupted data detected by QCS. Nevertheless, its capacity to induce a difference in terms of accuracy and statistical power for a given study can only be tested in a clinical context, with a real stratification or clustering question asked. Therefore, the objective of the second part paper is to evaluate the impact of QCS on clinical cohorts with a precise clinical question to demonstrate the clinical impact and the pertinence of the proposed approach.

**Acknowledgments**

The authors thank the subjects for their participation in this study. We thank Guillaume Y. Millet for his expertise in exercise physiology and involvement in this project. This work was partly supported by the LABEX PRIMES (ANR-11-LABX-0063), Siemens Healthineers and Jabrane Karkouri was supported by the European Union's Horizon 2020 research and innovation program under grant agreement No 801075.

**Data Availability Statement**

The data that support the findings of this study are available on request from the corresponding author. The data are not publicly available due to privacy or ethical restrictions.

**References:**

1. Enoka RM, Duchateau J. Translating Fatigue to Human Performance. *Medicine & Science in Sports & Exercise*. 2016;48(11):2228-2238. doi:10.1249/MSS.0000000000000929

2. Macintosh BR, Rassier DE. What Is Fatigue? *Can J Appl Physiol*. 2002;27(1):42-55. doi:10.1139/h02-003

3. Weiss K, Schär M, Panjrath GS, et al. Fatigability, Exercise Intolerance, and Abnormal Skeletal Muscle Energetics in Heart Failure. *Circ: Heart Failure*. 2017;10(7):e004129. doi:10.1161/CIRCHEARTFAILURE.117.004129

4. Meyerspeer M, Boesch C, Cameron D, et al. 31P magnetic resonance spectroscopy in skeletal muscle: Experts' consensus recommendations. *NMR in Biomedicine*. 2021;34(5). doi:10.1002/nbm.4246

5. Kemp GJ, Meyerspeer M, Moser E. Absolute quantification of phosphorus metabolite concentrations in human musclein vivo by 31P MRS: a quantitative review. *NMR Biomed*. 2007;20(6):555-565. doi:10.1002/nbm.1192

6. Liu Y, Gu Y, Yu X. Assessing tissue metabolism by 31P magnetic resonance spectroscopy and imaging: a methodology review. *Quant Imaging Med Surg*. 2017;7(6):707-716. doi:10.21037/qims.2017.11.03

7. Roussel M, Bendahan D, Mattei JP, Le Fur Y, Cozzone PJ. 31P Magnetic resonance spectroscopy study of phosphocreatine recovery kinetics in skeletal muscle: the issue of intersubject variability. *Biochimica et Biophysica Acta (BBA) - Bioenergetics*. 2000;1457(1-2):18-26. doi:10.1016/S0005-2728(99)00111-5

8. Ratkevicius A, Mizuno M, Povilonis E, Quistorff B. Energy metabolism of the gastrocnemius and soleus muscles during isometric voluntary and electrically induced contractions in man. *The Journal of Physiology*. 1998;507(2):593-602. doi:10.1111/j.1469-7793.1998.593bt.x

9. Layec G, Gifford JR, Trinity JD, et al. Accuracy and precision of quantitative 31P-MRS measurements of human skeletal muscle mitochondrial function. *American Journal of Physiology-Endocrinology and Metabolism*. 2016;311(2):E358-E366. doi:10.1152/ajpendo.00028.2016

10. Rzanny R, Stutzig N, Hiepe P, Gussew A, Thorhauer HA, Reichenbach JR. The reproducibility of different metabolic markers for muscle fiber type distributions investigated by functional 31P-MRS during dynamic exercise. *Zeitschrift für Medizinische Physik*. 2016;26(4):323-338. doi:10.1016/j.zemedi.2016.06.006

11. Edwards LM, Tyler DJ, Kemp GJ, et al. The Reproducibility of 31-Phosphorus MRS Measures of Muscle Energetics at 3 Tesla in Trained Men. Calbet JAL, ed. *PLoS ONE*. 2012;7(6):e37237. doi:10.1371/journal.pone.0037237

12. Layec G, Bringard A, Le Fur Y, et al. Reproducibility assessment of metabolic variables characterizing muscle energetics in Vivo: A $^{31}$P-MRS study: Muscle Metabolism Reliability. *Magn Reson Med*. 2009;62(4):840-854. doi:10.1002/mrm.22085

13. Šedivý P, Christina Kipfelsberger M, Dezortová M, et al. Dynamic $^{31}$P MR spectroscopy of plantar flexion: Influence of ergometer design, magnetic field strength (3 and 7 T), and RF-coil design:

Dynamic ³¹P MRS using different ergometers, MR-systems, and RF-coils. *Med Phys*. 2015;42(4):1678-1689. doi:10.1118/1.4914448

14. Luo Y, de Graaf RA, DelaBarre L, Tannús A, Garwood M. BISTRO: An outer-volume suppression method that tolerates RF field inhomogeneity: $B_1$-Insensitive Outer-Volume Suppression. *Magn Reson Med*. 2001;45(6):1095-1102. doi:10.1002/mrm.1144

15. Taylor DJ, Bore PJ, Styles P, Gadian DG, Radda GK. Bioenergetics of intact human muscle. A 31P nuclear magnetic resonance study. *Mol Biol Med*. 1983;1(1):77-94.

16. Valkovič L, Chmelík M, Krššák M. In-vivo 31P-MRS of skeletal muscle and liver: A way for non-invasive assessment of their metabolism. *Analytical Biochemistry*. 2017;529:193-215. doi:10.1016/j.ab.2017.01.018

17. Ratiney H, Sdika M, Coenradie Y, Cavassila S, Ormondt D van, Graveron-Demilly D. Time-domain semi-parametric estimation based on a metabolite basis set. *NMR Biomed*. 2005;18(1):1-13. doi:10.1002/nbm.895

18. Blümler P. MATLAB NMR-Library. https://www.blogs.uni-mainz.de/fb08-physics-halbach-magnets/software/

19. Sundberg CW, Prost RW, Fitts RH, Hunter SK. Bioenergetic basis for the increased fatigability with ageing. *J Physiol*. 2019;597(19):4943-4957. doi:10.1113/JP277803

20. Slade JM, Abela GS, Rozman M, et al. The impact of statin therapy and aerobic exercise training on skeletal muscle and whole-body aerobic capacity. *American Heart Journal Plus: Cardiology Research and Practice*. 2021;5:100028. doi:10.1016/j.ahjo.2021.100028

21. Forbes SC, Slade JM, Francis RM, Meyer RA. Comparison of oxidative capacity among leg muscles in humans using gated ³¹P 2-D chemical shift imaging: PHOSPHOCREATINE RECOVERY KINETICS. *NMR Biomed*. 2009;22(10):1063-1071. doi:10.1002/nbm.1413

22. Forbes SC, Slade JM, Meyer RA. Short-term high-intensity interval training improves phosphocreatine recovery kinetics following moderate-intensity exercise in humans. *Appl Physiol Nutr Metab*. 2008;33(6):1124-1131. doi:10.1139/H08-099

23. Forbes SC, Paganini AT, Slade JM, Towse TF, Meyer RA. Phosphocreatine recovery kinetics following low- and high-intensity exercise in human triceps surae and rat posterior hindlimb muscles. *Am J Physiol Regul Integr Comp Physiol*. 2009;296(1):R161-R170. doi:10.1152/ajpregu.90704.2008

24. Brown R, Sharafi A, Slade JM, et al. Lower extremity MRI following 10-week supervised exercise intervention in patients with diabetic peripheral neuropathy. *BMJ Open Diab Res Care*. 2021;9(1):e002312. doi:10.1136/bmjdrc-2021-002312

25. Hurley DM, Williams ER, Cross JM, et al. Aerobic Exercise Improves Microvascular Function in Older Adults. *Medicine & Science in Sports & Exercise*. 2019;51(4):773-781. doi:10.1249/MSS.0000000000001854

26. Kreis R, Boer V, Choi I, et al. Terminology and concepts for the characterization of in vivo MR spectroscopy methods and MR spectra: Background and experts' consensus recommendations. *NMR in Biomedicine*. 2021;34(5). doi:10.1002/nbm.4347

27. Iotti S, Gottardi G, Clementi V, Barbiroli B. The mono-exponential pattern of phosphocreatine recovery after muscle exercise is a particular case of a more complex behaviour. *Biochimica et Biophysica Acta (BBA) - Bioenergetics*. 2004;1608(2-3):131-139. doi:10.1016/j.bbabio.2003.11.003

28. Iotti S, Borsari M, Bendahan D. Oscillations in energy metabolism. *Biochimica et Biophysica Acta (BBA) - Bioenergetics*. 2010;1797(8):1353-1361. doi:10.1016/j.bbabio.2010.02.019

29. Jill M.Slade, Towse TF, DeLano MC, Wiseman RW, Meyer RA. A gated 31P NMR method for the estimation of phosphocreatine recovery time and contractile ATP cost in human muscle. *NMR Biomed*. 2006;19(5):573-580. doi:10.1002/nbm.1037

30. Gregory CM, Vandenborne K, Dudley GA. Metabolic enzymes and phenotypic expression among human locomotor muscles. *Muscle Nerve*. 2001;24(3):387-393. doi:10.1002/1097-4598(200103)24:3<387::AID-MUS1010>3.0.CO;2-M

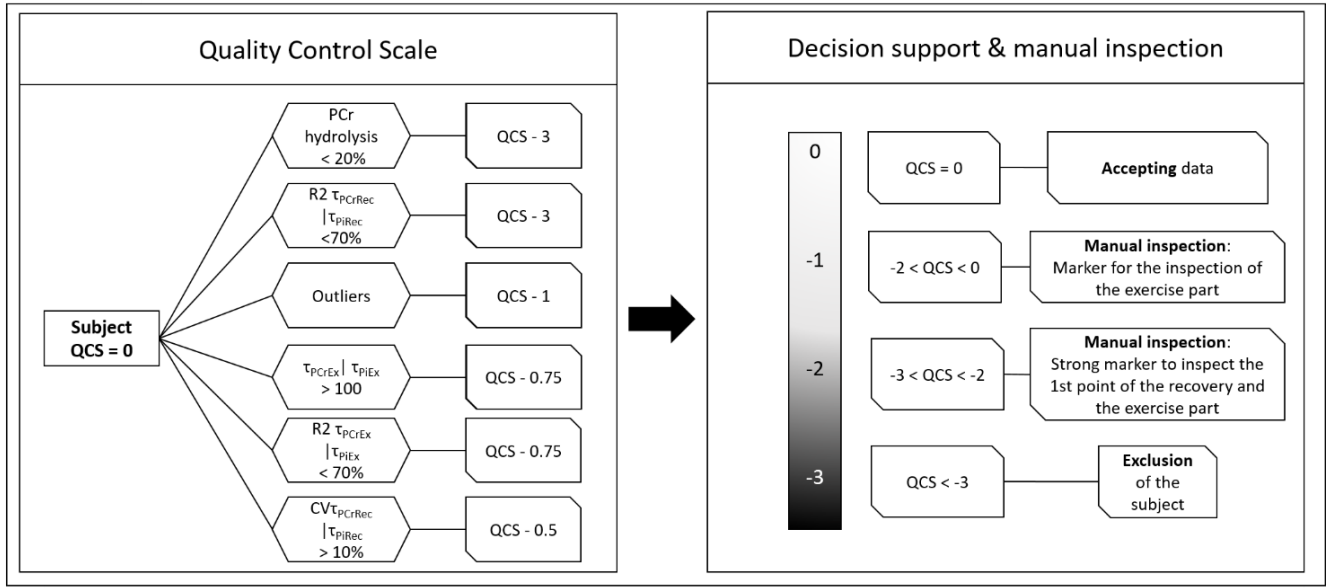

Figure 1: The rating scale reflecting the quality of the data acquisition and the estimated parameters. Left: for a given subject, each criterion violation leads to a greater or lesser decrease in the score, depending on the importance given to the criteria. Right: Depending on the score, the operator can exclude or include the exercise part in the analysis and validate the first point of the recovery.

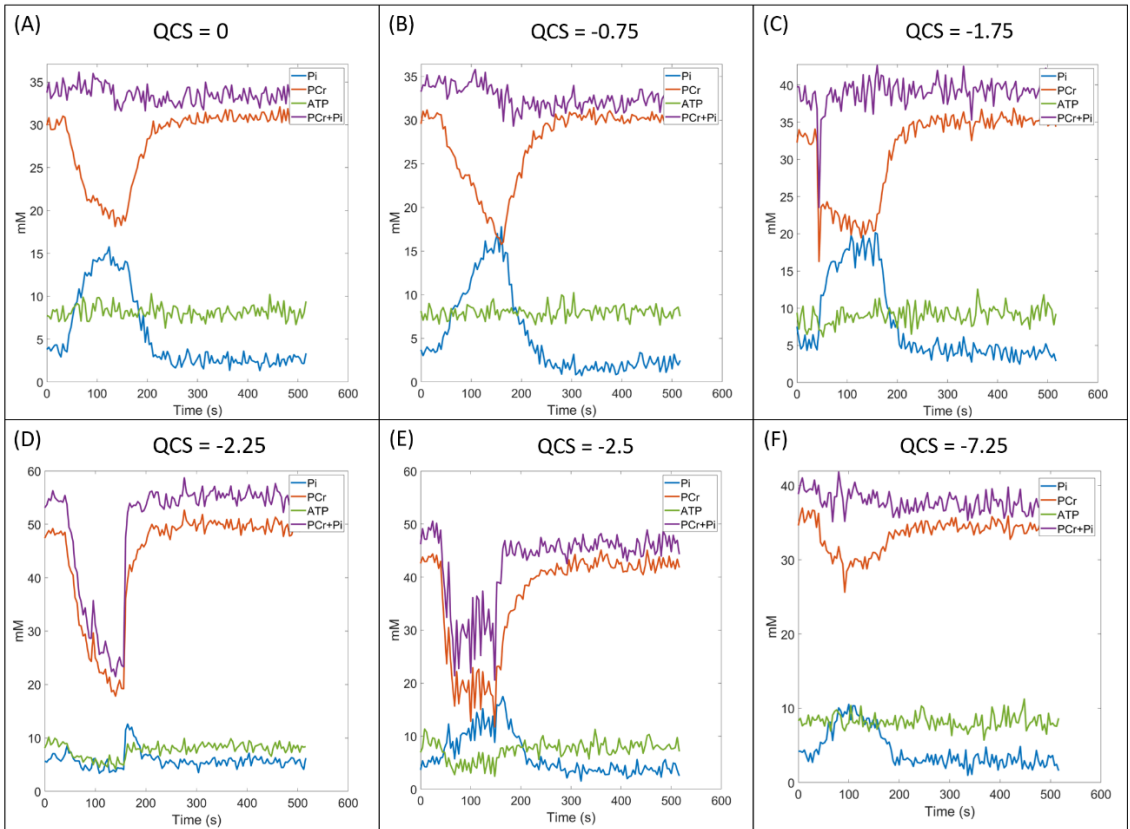

Figure 2: Raw data from representative subjects showing different patterns highlighted by the Quality Control Score of dynamic datasets. (A) QCS=0, all data accepted; (B) QCS=-0.75, with τPCrEx| τPiEx > 100s: exercise data rejected; (C) QCS=-1.75 with outliers criteria and the R2 τPCrEx|τPiEx < 70%: exercise data rejected; (D) QCS=-2.25, with outliers criteria: exercise data rejected and reselection of the first point for fitting the metabolite recoveries; (E) QCS=-2.5, with outliers criteria: exercise data rejected; (F) QCS=-7.25: PCr hydrolysis<20% and R2 τPCrEx|τPiEx< 70%: all data rejected.

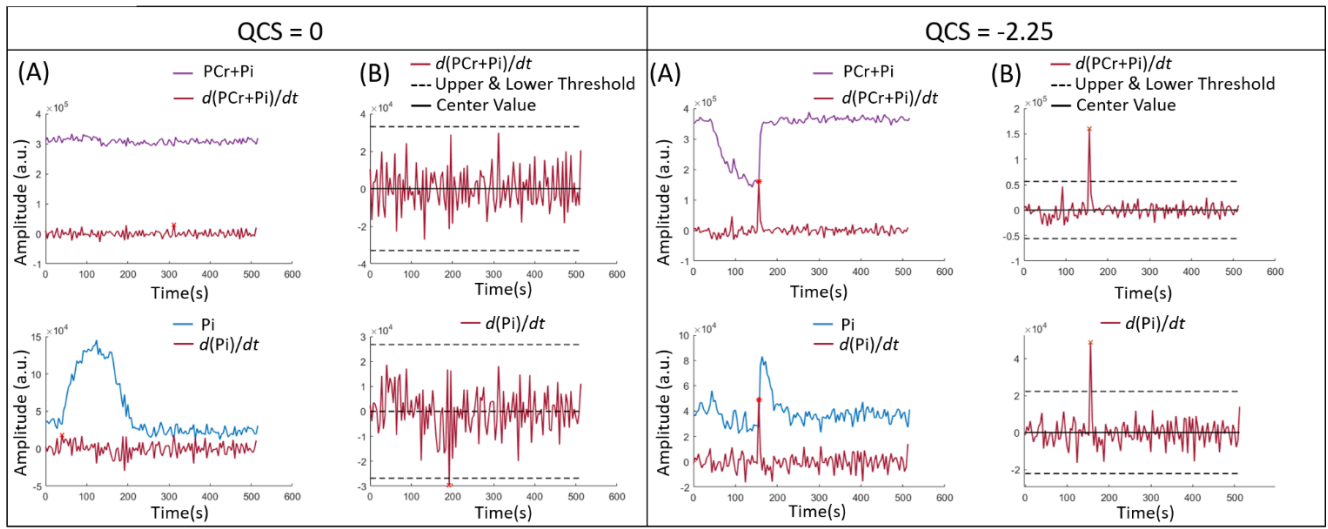

Figure 3: Amplitude of metabolites during the protocol and their derived value in the case of a COVID19 patient with acquisition errors and control. Left: subject with QCS = 0. Right : subject with QCS=-2.25. (A), Amplitude, derived value and maximum value of (PCr+Pi) and Pi, (B), Derived value and associated threshold for outlier detection. The positive values of the outliers are retained and considered for the choice of the first fit point. For the subjects with CQS=-2.25, the outlier is at point 156s, the best point to consider will be at 160s.

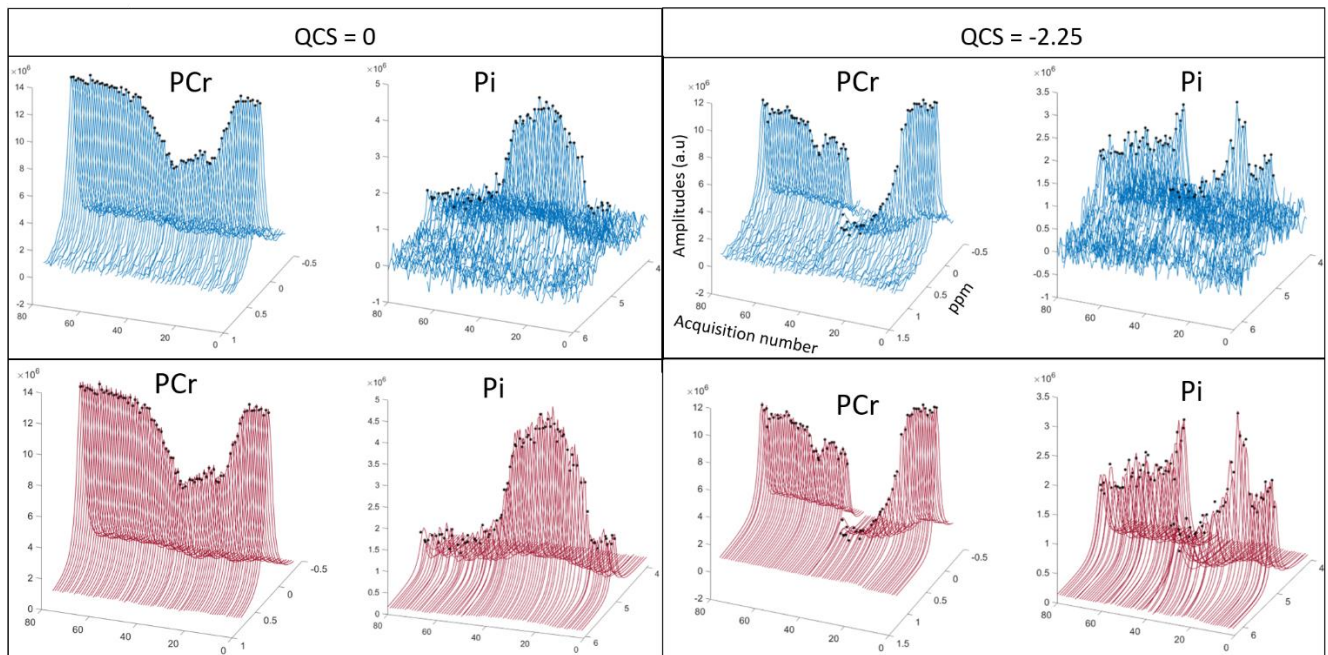

Figure 4: Stack plots of PCr and Pi acquired during the dynamic protocol (rest, exercise, recovery) for a subject without acquisition errors (left) and for a subject with errors (right). Top are the raw spectra; bottom are the estimated QUEST spectra.
23

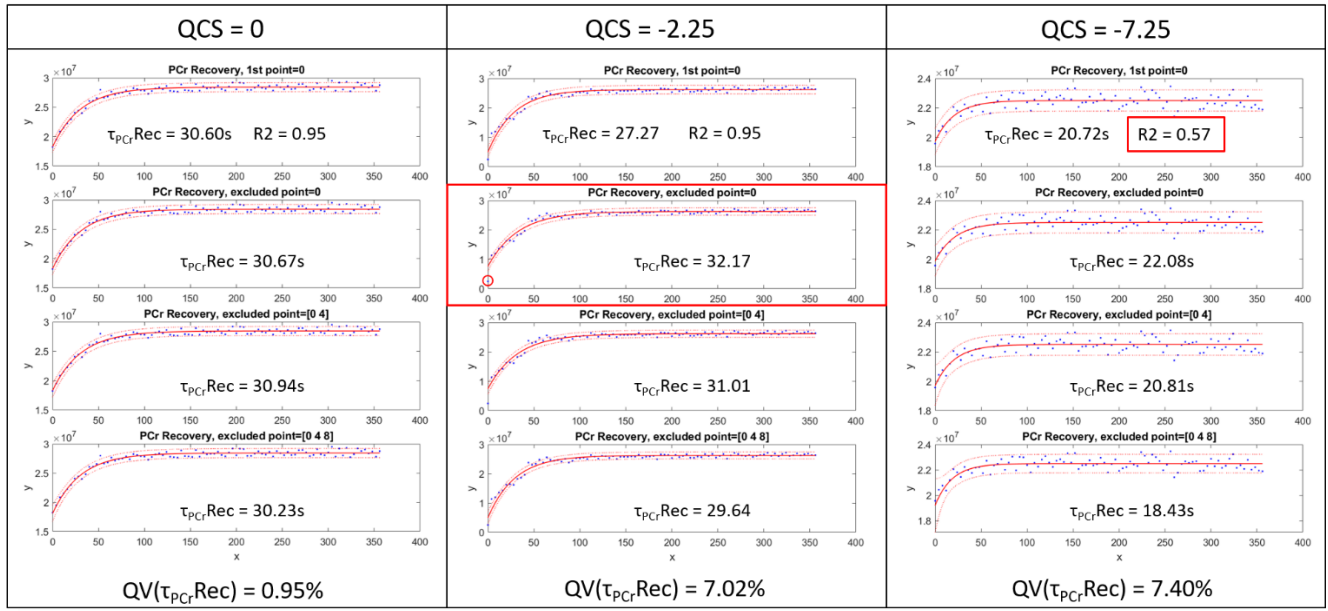

Figure 5: PCr fitting for $\tau_{PCr}$Rec quality control assessment. The left shows a subject with good quality data with no data points removed. The middle shows a subject with moderately reduced QCS. The subject presents with acquisition errors throughout the exercise, which affects the first point of the recovery. After inspection, the first point of the fitting procedure is excluded and the τ-increases from 27.3s to 32.2s. The right shows a subject with very low-quality data and low QCS. The subject has an R2 coefficient lower than the acceptable level (<70) and τPCr for recovery is rejected and thereby the subject excluded for this outcome measure.

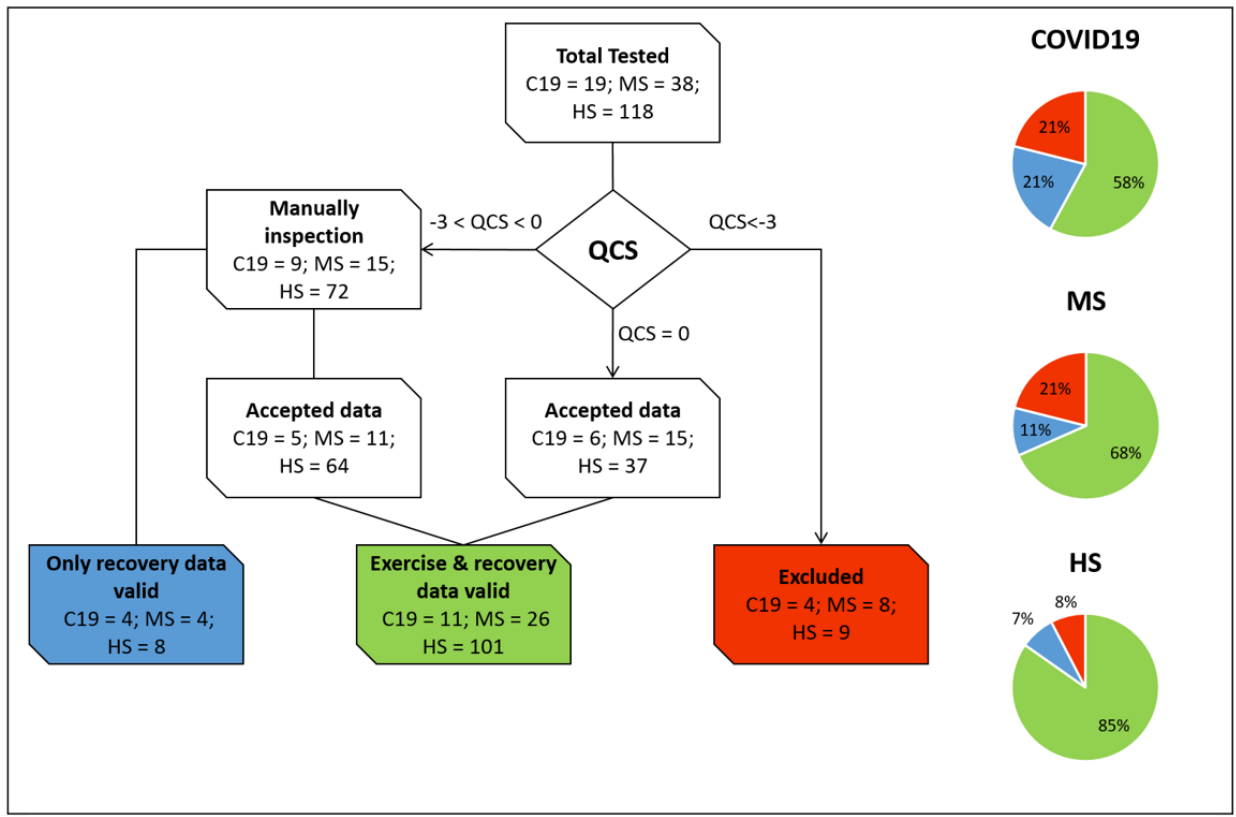

Figure 6: Flowchart of the application of the quality control score. The flowchart boxes show the resulting subject numbers with acceptable data, excluded data and those requiring manually inspection for ultimate decision making on data acceptance. Pie charts display the results of the QCS application as percentages across the clinical and healthy groups. C19=COVID19 patients; MS=Multiple Sclerosis patients; HS= Healthy Control Subjects.